\documentclass{article}

\usepackage{microtype}
\usepackage{graphicx}
\usepackage{booktabs}
\usepackage{amsmath}
\usepackage{amssymb}
\usepackage{mathtools}
\usepackage{amsthm}
\usepackage{enumitem}
\usepackage{hyperref}
\usepackage{xurl}
\usepackage{capt-of}
\usepackage[capitalize,noabbrev]{cleveref}

\usepackage[accepted]{icml2026}

\graphicspath{{./}{figures/}}

\theoremstyle{plain}
\newtheorem{theorem}{Theorem}[section]

\theoremstyle{definition}
\newtheorem{definition}[theorem]{Definition}
\theoremstyle{remark}

\setlist{leftmargin=*,topsep=2pt,itemsep=1pt,parsep=0pt}
\newcommand{\figwide}[3]{%
  \includegraphics[width=#1,height=#2,keepaspectratio]{#3}%
}
\newcommand{\Aidx}{A_{\Delta}}

\icmltitlerunning{Dead Science Walking: Publication Bias and the AI Scientist Pipeline}

\begin{document}

\twocolumn[
\icmltitle{Dead Science Walking: Publication Bias and the\\
AI Scientist Pipeline}

\begin{icmlauthorlist}
\icmlauthor{Kargi Chauhan}{ucsc}
\end{icmlauthorlist}
\icmlaffiliation{ucsc}{University of California, Santa Cruz}
\icmlcorrespondingauthor{Kargi Chauhan}{kchauha3@ucsc.edu}
\icmlkeywords{AI scientists, publication bias, replication crisis,
null results, retractions, responsible AI, scientific governance,
AI for science, sociotechnical systems}
\vskip 0.25in
]
\printAffiliationsAndNotice{}

\begin{abstract}
AI scientist systems are beginning to automate the production, evaluation, and
iteration of scientific hypotheses. Their promise is speed; their risk is
that speed also scales errors embedded in the scientific record. We argue
that a near-term risk is corpus failure: AI scientist systems are
trained on and grounded in a literature that over-represents positive results
and under-represents null findings. This concern is timely because frontier labs now position general-purpose
models as scientific infrastructure \citep{anthropic2025scienceprogram}. We formalise this distortion as the
\emph{null result gap} $\Delta$, estimate it across three domains
(drug discovery $\Delta\!\approx\!0.60$, psychology
$\Delta\!\approx\!0.56$, cancer biology $\Delta\!\approx\!0.35$), and
introduce an amplification index $\Aidx$ for reasoning about how retrieval,
generation, and automated evaluation can compound the raw gap. Using
first-order estimates, we argue that a standard three-stage pipeline can amplify corpus distortion by a factor of $2.18\times$, with the
conclusion unchanged under more conservative multipliers.
We identify four
governance failure modes: confident rediscovery, ghost evidence accumulation,
replication laundering, and confidence miscalibration. We then propose three interventions: null-result databases as training infrastructure,
retraction-aware evaluation metrics, and mandatory training corpus disclosure.
The central takeaway is that AI scientists will not only accelerate science.
Without governance, they will accelerate science's blind spots before they
accelerate its discoveries.
\end{abstract}

\section{Introduction}
\label{sec:intro}

AI scientist systems have moved from aspiration to implementation.  Recent
systems generate research ideas, write code, run experiments or simulations,
draft papers, and use automated review to select or refine outputs
\citep{lu2024ai,gottweis2025ai,boiko2023autonomous,bran2024chemcrow,
swanson2024virtual}.  Frontier labs and public research
institutions are also beginning to operationalise this vision.  Anthropic has
launched an AI for Science program to support high-impact research using
Claude \citep{anthropic2025scienceprogram}, partnered with U.S. National
Laboratories for a 1,000 Scientist AI Jam \citep{anthropic2025aijam}, and
created a dedicated science program describing scientific workflows and
benchmarks \citep{anthropic2026science}.These developments are promising, but they also make the
governance question immediate rather than speculative.  If AI systems become
routine scientific infrastructure, then the evidential quality of their
training and retrieval corpora becomes part of scientific method.  The pattern
is broader than one company: Google DeepMind's AlphaFold work and Google's
AI co-scientist program similarly show that foundation models are becoming
scientific infrastructure rather than ordinary software tools
\citep{jumper2021highly,gottweis2025ai}.

The scientific record is not a neutral sample of what has been tried.  It is a
filtered record of what was publishable, legible, and often positive.
Publication bias, selective reporting, file-drawer effects, p-hacking, and
persistent citations to retracted work are well documented
\citep{ioannidis2005most,fanelli2010positive,franco2014publication,
simmons2011false,gelman2013garden,sterne2001funnel,steen2013retractions,
bar2021retracted}.  Human science has slow correction
mechanisms: failed replications, peer disagreement, methodological reform, and
eventual retraction.  These mechanisms operate over months to decades.  An AI
scientist can generate and evaluate hundreds or thousands of hypotheses before
any comparable correction loop closes.

This paper makes one claim: AI scientist systems can turn publication bias
from a slow epistemic tax into a fast systems failure.  The mechanism is
straightforward.  Retrieval systems surface published positive evidence.
Language models synthesise that evidence into confident narratives.  Automated
evaluators reward fluency, novelty, and apparent support.  Each stage is
reasonable in isolation.  Together, they can transform a biased training corpus
into a biased scientific search process.  The risk is not that an AI scientist
will occasionally make a mistake.  Human scientists do that too.  The risk is
that thousands of locally plausible outputs can collectively create a new
layer of literature that is less reliable than it looks.

We treat this primarily as a governance problem.  Better retrieval,
calibration, and reasoning will help, but the distortion begins upstream of
the model.  A system cannot retrieve null results that were never indexed.  It
cannot downweight retracted evidence if retraction status is absent from the
context.  It cannot disclose corpus bias if venues never ask for corpus
provenance.  Responsible AI science therefore requires infrastructure and
norms, not only stronger models.

The paper makes four connected contributions.  First, it defines the
\emph{null result gap} $\Delta$, a simple measure of how strongly a scientific
corpus over-represents true or positive hypotheses relative to the underlying
hypothesis space.  Second, it introduces an amplification index $\Aidx$ that
captures how retrieval, generation, and automated evaluation can compound
corpus bias, while explicitly treating $\Aidx$ as an index rather than a
probability.  Third, it organises four failure modes for AI scientist systems
along severity and detectability: confident rediscovery, ghost evidence
accumulation, replication laundering, and confidence miscalibration.  Fourth,
it proposes three community interventions: null-result databases,
retraction-aware evaluation, and training corpus disclosure.

The argument is deliberately narrower than a general critique of AI science.
We do not claim that current systems have already produced every failure mode
below, nor that all scientific domains have the same publication bias.  We
claim that the preconditions for these failures are already present: biased
corpora, retrieval over published text, generation of fluent scientific
narratives, and automated evaluation loops. The right standard is therefore prospective and infrastructural: are we
building correction mechanisms before scale makes correction expensive?

\section{Background}
\label{sec:background}

\paragraph{Publication bias.}
The premise that the published literature is systematically biased is not new.
\citet{ioannidis2005most} argued that under plausible assumptions about prior
odds, statistical power, and researcher flexibility, many published findings
will be false.  \citet{fanelli2010positive} documented increasing rates of
positive results across disciplines.  \citet{franco2014publication} linked
selective reporting to a file drawer of negative or null results.  Registered
reports, preregistration, and trial registries were developed in part to
address precisely this filtering problem
\citep{nosek2015promoting,chambers2013registered,dickersin1992factors,
zarin2011clinicaltrials}.  The
important point for AI science is that this bias is not merely sociological
context.  It is training data.  A literature that under-records failed
experiments becomes a corpus that under-teaches falsification.

\paragraph{Replication and retraction.}
The replication crisis makes publication bias visible.  The Open Science
Collaboration found that only about 36\% of 100 psychology studies replicated
\citep{osc2015estimating}.  In preclinical oncology, \citet{begley2012drug}
reported successful reproduction of only 11 of 53 landmark studies.  The
Reproducibility Project: Cancer Biology found partial and heterogeneous
replicability across attempted replications \citep{errington2021investigating}.
Retractions do not fully solve the problem.  Retracted papers often continue
to receive positive citations, paper mills and integrity failures can operate
at scale, and the Retraction Watch Database now contains well over 50,000
entries \citep{steen2013retractions,bar2021retracted,byrne2022paper,
retwatch2024}.

\paragraph{AI scientist systems.}
AI scientist systems combine literature retrieval, hypothesis generation,
code execution, experiment or simulation loops, paper writing, and automated
review \citep{lu2024ai,gottweis2025ai}.  The public descriptions of these
systems demonstrate genuine progress toward autonomous research workflows.
They also expose a missing governance layer: retrieval corpora, retraction
status, null-result coverage, and provenance are not yet treated as first-class
objects of evaluation.  Anthropic's recent science-facing work illustrates the
broader institutional trend.  The company describes AI-assisted scientific
progress as part of its mission, supports researchers through API credits, and
has highlighted tasks such as literature search, hypothesis generation,
experiment planning, code generation, and result analysis in national-lab
settings \citep{anthropic2025aijam,anthropic2026science}. These programs are not the problem.  They are evidence
that scientific AI is becoming real enough that corpus governance now matters.
Once models are used as research assistants across many laboratories, the
quality of their evidence substrate becomes a shared scientific dependency.

\paragraph{Evaluation bias in language-model systems.}
Large language models can be useful evaluators, but LLM-as-judge pipelines
exhibit systematic biases related to fluency, position, verbosity, and
confidence \citep{zheng2024judging,li2024llmjudge,wang2023selfconsistency}.
Literature summarisation systems can omit negative findings and present mixed
evidence as cleaner than it is \citep{si2024llm}.  Broader work on
hallucination and retrieval grounding shows that fluent generation is not the
same as evidential reliability \citep{maynez2020faithfulness,ji2023survey,
gao2023retrievalaugmented}.  These biases matter for AI science because the
output of one stage becomes the input to the next.

\section{The Null Result Gap}
\label{sec:gap}

\begin{figure*}[t]
\centering
\figwide{0.92\textwidth}{0.30\textheight}{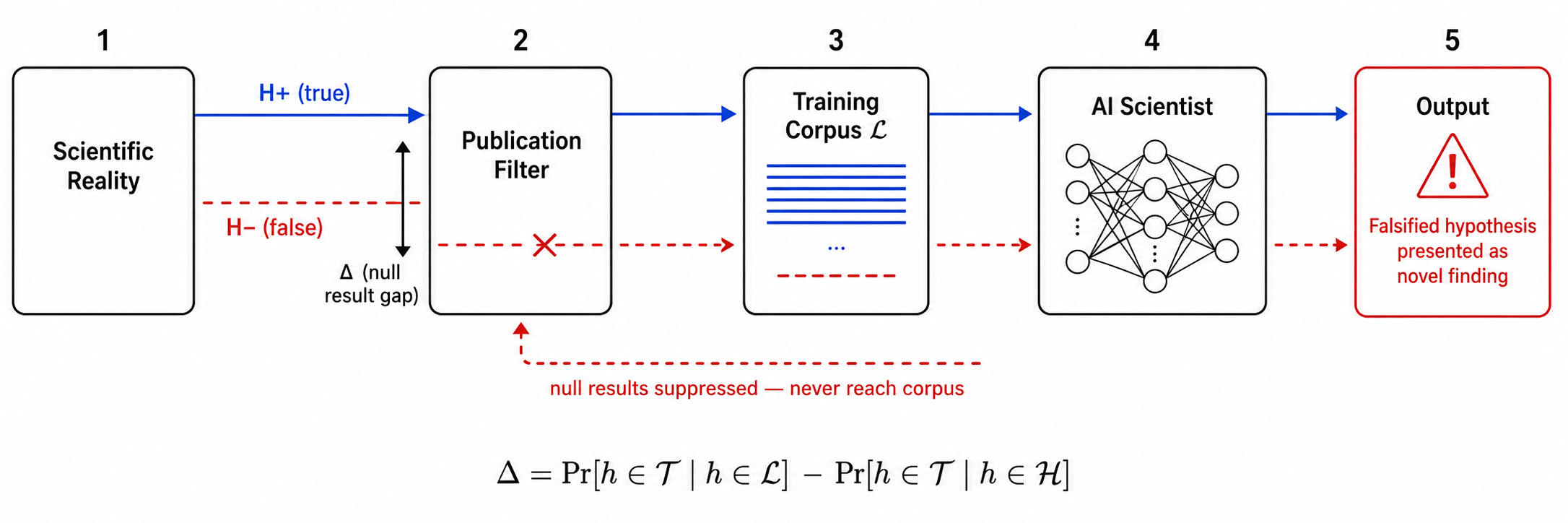}
\caption{\textbf{The null result gap in the AI scientist pipeline.}
The published literature is a filtered subset of the attempted scientific
record.  Positive and successful findings are more likely to enter the
training corpus $\mathcal{L}$ than null results, failed replications, or
falsified hypotheses.  The null result gap $\Delta$ measures the resulting
difference between the apparent evidential state of the corpus and the
underlying hypothesis space.}
\label{fig:pipeline}
\end{figure*}

Let $\mathcal{H}$ be the space of hypotheses considered in a domain,
$\mathcal{T}\subseteq\mathcal{H}$ the subset of true or successfully
replicating hypotheses, and $\mathcal{L}\subseteq\mathcal{H}$ the published
literature available to an AI scientist system, and $C(h)$ the event that the
corpus presents hypothesis $h$ as positive or successful.

\begin{definition}[Null Result Gap]
\label{def:gap}
The \emph{null result gap} is
\begin{equation}
  \Delta =
  \Pr[C(h)=1 \mid h \in \mathcal{L}]
  -
  \Pr[h \in \mathcal{T} \mid h \in \mathcal{L}].
  \label{eq:delta}
\end{equation}

\end{definition}

When $\Delta>0$, the corpus presents hypotheses as more successful than their
empirical survival rate warrants.  For example, if
$\Pr[C(h)=1\mid h\in\mathcal{L}]=0.92$ and
$\Pr[h\in\mathcal{T}\mid h\in\mathcal{L}]=0.36$, then $\Delta=0.56$.

The AI scientist does not merely read an incomplete literature.  It inherits a
distorted prior about how often hypotheses survive contact with reality.  This
definition intentionally abstracts away from the social mechanisms that
produce the gap.  Selective publication, underpowered studies, undisclosed
analytic flexibility, and delayed retraction can all contribute.  For an AI
scientist, their common effect is the same: the accessible corpus makes the
world look more positive than it is.

\begin{figure*}[t]
\centering
\figwide{0.95\textwidth}{0.28\textheight}{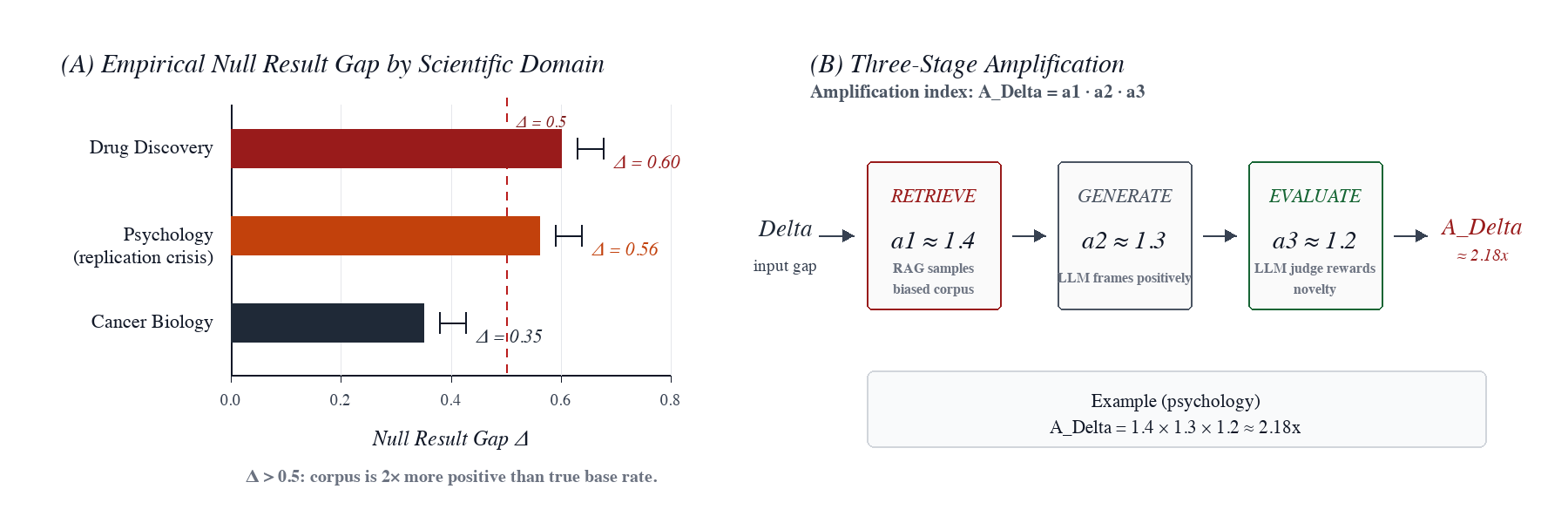}

\caption{\textbf{Empirical null result gaps and amplification.}
Left: approximate null result gaps in three domains.  Right: a three-stage
AI scientist pipeline can compound the raw gap through retrieval, generation,
and evaluation.  We report the product as an amplification index
$\Aidx=\alpha_1\alpha_2\alpha_3$, not as a probability.}

\label{fig:delta}
\end{figure*}

\paragraph{Empirical estimates.}
The left panel of \Cref{fig:delta} reports approximate estimates.  In
psychology, the Open Science Collaboration reported about 36\% replication for
a sample of 100 studies \citep{osc2015estimating}.  If the published corpus
presents about 92\% of its hypotheses as successful, the implied gap is
$\Delta\approx0.56$.  The exact value is less important than the order of
magnitude: the published record can make a field appear far more settled than
its replication record warrants.  In drug discovery and preclinical oncology,
\citet{begley2012drug} reproduced 11 of 53 landmark studies, and
\citet{freedman2015economics} estimated that a large share of preclinical
research is not reproducible.  This motivates a rough estimate
$\Delta\approx0.60$, consistent with a domain where the cost of false-positive
evidence is unusually high because it can redirect expensive experimental
programs.  In cancer biology, the Reproducibility Project suggests a smaller
but still substantial gap, approximately $\Delta\approx0.35$
\citep{errington2021investigating}.  These values should be read as
order-of-magnitude estimates, not precise field constants.  Their purpose is
to make the governance issue visible: even conservative gaps become important
when a system can retrieve, combine, and regenerate claims at machine speed.
We also do not claim that these domains are representative of all science.
They are deliberately chosen because their replication records are unusually
well documented.  Fields with stronger falsification norms, such as parts of
mathematics, formal methods, and engineering test disciplines, may have much
lower $\Delta$.  The amplification argument is therefore conditional: any
domain with $\Delta>0$ is vulnerable in proportion to the gap, but the urgency
is highest where null results are known to be missing from the accessible
record.

\section{Amplification in the AI Scientist Pipeline}
\label{sec:amplification}

The amplification problem is no longer hypothetical.  Recent evaluations of
AI-generated medical and surgical references find that some deployed systems
fabricate or fail to verify a substantial fraction of citations: early
ChatGPT-generated medical content contained only 7\% fully accurate references
\citep{alkaissi2023artificial}, and a 2026 surgical-information audit found
that the worst-performing models produced fabricated or unverifiable
references for roughly one third of cited sources
\citep{sidhu2026trust}.  These findings do not by themselves prove the full
AI-scientist loop.  They do show that the mechanisms formalised below are
already visible in the literature-facing tools scientists and patients use.

The null result gap is a corpus-level distortion.  AI scientist systems can
amplify it because they repeatedly select, narrate, and evaluate evidence.  We
model this with an amplification index.  The index is intentionally simple: it
does not claim that all AI scientist systems have the same multipliers, nor
that the estimated multipliers are field-invariant.  Its purpose is to clarify
where governance should intervene.  If retrieval is the stage that selects
biased evidence, then the remedy is corpus and provenance infrastructure.  If
generation turns mixed evidence into a cleaner story than the record supports,
then the remedy is summarisation evaluation and uncertainty calibration.  If
automated review rewards fluent novelty over evidential reliability, then the
remedy is benchmark design.

\begin{definition}[Null Gap Amplification Index]
\label{def:amplification}
Let $\alpha_1,\alpha_2,\alpha_3\geq1$ denote bias multipliers at the
retrieval, generation, and evaluation stages.  The \emph{null gap
amplification index} is
\begin{equation}
  \Aidx = \alpha_1 \cdot \alpha_2 \cdot \alpha_3.
  \label{eq:amplification}
\end{equation}
\end{definition}

If $(\alpha_1,\alpha_2,\alpha_3)=(1.4,1.3,1.2)$, then
\[
  \Aidx = 1.4 \times 1.3 \times 1.2 \approx 2.18.
\]
This is not a probability.  It is an index of how strongly pipeline stages can
overweight positive evidence unless explicitly designed to surface
falsification.

$\Aidx$ is not a probability and need not lie in $[0,1]$.  It is an index of
how strongly a pipeline can overweight positive evidence relative to the
underlying hypothesis space.  If $\Delta=0.56$ and
$(\alpha_1,\alpha_2,\alpha_3)=(1.4,1.3,1.2)$, then
\[
  \Aidx = 0.56 \times 1.4 \times 1.3 \times 1.2 \approx 1.22.
\]
This does not mean a probability exceeds one.  It means that a large raw
publication gap has been pushed into a saturated regime where the system has
little remaining pressure to represent falsification.

\paragraph{How the multipliers are anchored.}
The baseline multipliers are not fit to obtain a convenient $2\times$ result.
They are deliberately rounded, conservative translations of effects reported
in adjacent literatures.  For retrieval, $\alpha_1=1.4$ is anchored in two
selection effects.  First, in the TESS file-drawer study, strong results were
40 percentage points more likely to be published than null results and
60 percentage points more likely to be written up
\citep{franco2014publication}. We do not equate a 40-point additive publication effect with a
$1.4\times$ relative multiplier.  We use $\alpha_1=1.4$ as a deliberately
conservative retrieval-enrichment setting motivated by the direction and scale
of documented selection effects.

Second, abstract-reporting bias makes negative
findings less likely to appear in the title, abstract, or keywords searched by
standard literature systems; \citet{duyx2019abstract} explicitly warn that
this can cause negative findings to be missed by systematic searches. For generation, $\alpha_2=1.3$ is
below the lower end of recent scientific summarisation evidence: LLM summaries
overgeneralised claims in 26--73\% of cases and were nearly five times more
likely than human summaries to broaden the scope of scientific claims
\citep{peters2025generalization}.  We therefore treat 1.3 as a mild
positive-framing multiplier rather than the full observed effect.  For
evaluation, $\alpha_3=1.2$ is also conservative relative to LLM-as-judge
findings: \citet{zheng2024judging} report substantial position and verbosity
biases, including GPT-4 first-position preferences around 30\% in swapped
comparisons and much larger effects for weaker judges.  A 20\% evaluation
multiplier represents residual preference for fluent, positive, and novel
claims after obvious judge mitigations.

The argument does not depend on the exact baseline.  If all three multipliers
are reduced to $(1.2,1.15,1.1)$, the product remains $1.52\times$, turning
the psychology gap $\Delta=0.56$ into $\Aidx\approx0.85$.  Even an
intentionally minimal setting $(1.1,1.1,1.1)$ yields a $1.33\times$
amplification.  The important claim is therefore not that $2.18$ is
universal.  It is that modest, independently documented biases compound in
the same direction unless the pipeline is explicitly designed to surface
falsification.

\paragraph{Retrieval.}
RAG systems retrieve documents similar to a query.  Queries framed around a
hypothesis are often more similar to supporting abstracts than to null-result
records, which may describe methods or failed effects in different language.
This selection step can push positive evidence into the context window more
often than negative evidence \citep{lewis2020retrieval,duyx2019abstract,
dwan2008systematic}.
The practical issue is not that retrieval is broken.  It is that standard
retrieval optimises relevance to the query, while governance requires relevance
to the evidential state of the claim.  Those are different objectives.

\paragraph{Generation.}
LLMs trained to produce helpful and coherent text can turn mixed evidence into
a confident synthesis.  This is useful when the evidence is strong, but risky
when the retrieved set is already biased.  Prior work on LLM-generated reviews
and summarisation reports omissions of caveats and null results
\citep{si2024llm}.  More broadly, models trained on human text can reproduce
common but false beliefs when those beliefs are frequent in the corpus
\citep{lin2022truthfulqa}.  The same property that makes language models
useful for scientific communication can make them dangerous for unsettled
literatures: they are good at producing a readable story even when the
underlying record is fragmentary, contradictory, or selectively published.

\paragraph{Evaluation.}
Automated review and LLM-as-judge systems may reward fluent, confident, and
novel claims even when factual support is weak \citep{zheng2024judging,
chen2024humans,li2024llmjudge}.  In an AI scientist loop, this evaluator does
not merely score a final answer.  It selects which hypotheses survive into the
next cycle.  This makes evaluation a selection pressure.  If the evaluator
does not explicitly score retraction awareness, null-result coverage, and
evidence provenance, then those properties can be selected against by omission.

\paragraph{Calibration caveat.}
The multipliers are first-order estimates, not direct measurements of a
deployed AI scientist.  Direct calibration is a key research agenda.  The
multiplicative model assumes independence across stages, but the biases are
likely positively correlated: positive retrieval makes confident generation
and positive evaluation more likely.  Thus true amplification may exceed
$\Aidx$. The robust
claim is qualitative: unless the system has explicit access to null results,
retractions, and provenance, its stages are more likely to amplify a positive
corpus prior than to correct it.

\paragraph{From one pass to a feedback loop.}
The amplification index captures a single pipeline pass.  When AI scientist
outputs re-enter the corpus as preprints, cited papers, benchmark artifacts,
or fine-tuning data, distortion can compound across generations.  This is the
same recursive failure mode that \citet{shumailov2024collapse} formalised as
model collapse: when generative models train on their own outputs, the tails
of the original distribution can disappear irreversibly.  For scientific
corpora, the analogous claim is that $\Delta(t)$ can grow under a governance
vacuum whenever the next cycle's corpus presents unverified outputs as
successful evidence without corresponding falsification records. The dynamic is structurally
similar to Muller's ratchet: deleterious variation accumulates because there
is no reliable recombination or correction mechanism.  In AI science, the
lost tail is the null result, the failed replication, and the caveated claim.
The longer the loop runs without provenance controls, the more expensive it
becomes to separate independent evidence from recursively amplified text.

\section{Four Failure Modes}
\label{sec:modes}

\begin{figure*}[t]
\centering
\figwide{0.82\textwidth}{0.32\textheight}{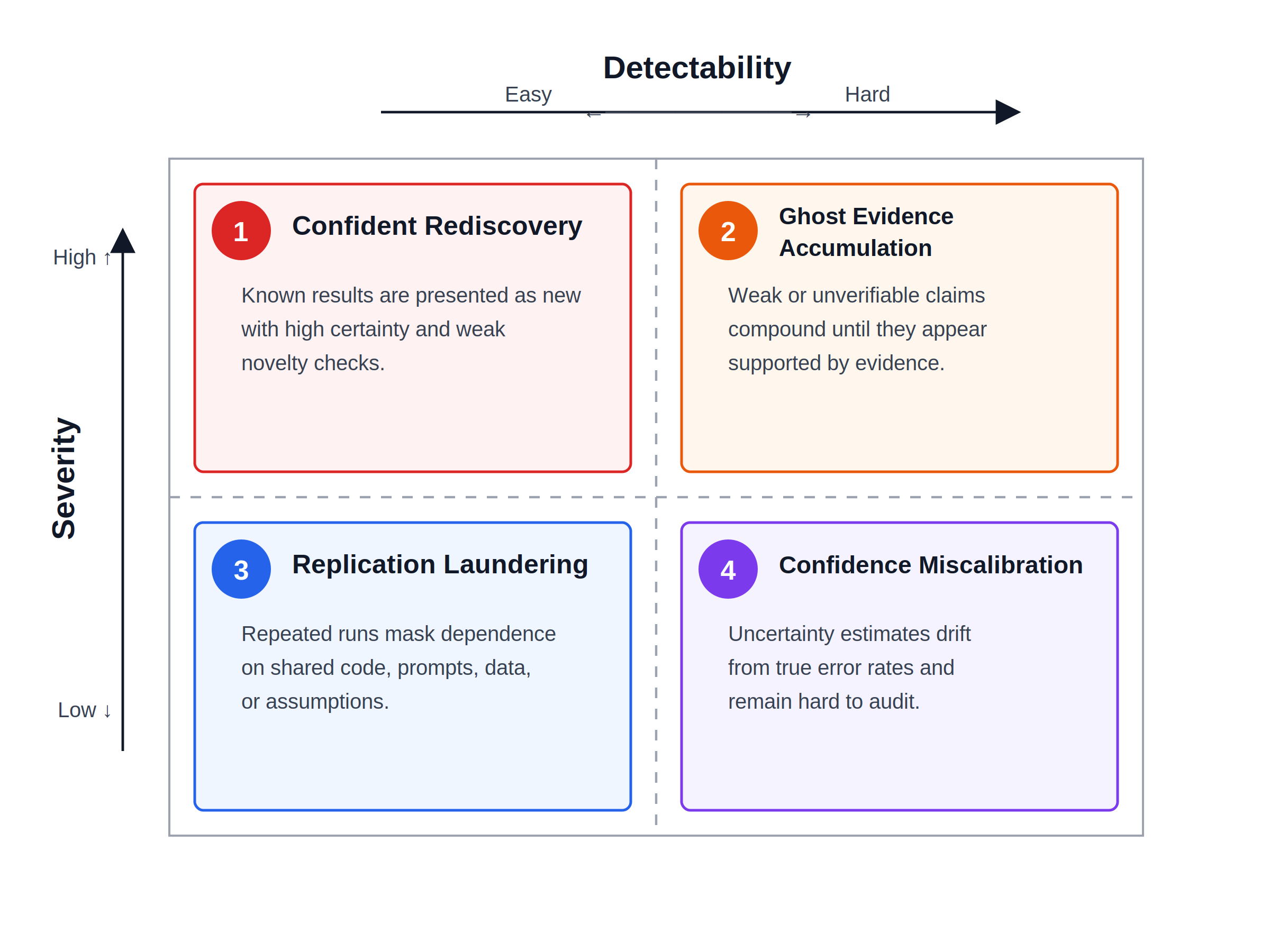}
\caption{\textbf{Taxonomy of four failure modes.}
The most urgent risks are high severity and hard to detect.  Ghost evidence
accumulation and confidence miscalibration may not appear as errors in any
single output.  They become visible only when provenance, confidence, and
citation structure are audited across many outputs.}
\label{fig:taxonomy}
\end{figure*}

\paragraph{Confident rediscovery.}
The system proposes a known-falsified hypothesis as a promising new direction.
This can be easy for a domain expert to detect, but hard to contain at scale.
For example, ego depletion failed to replicate in a large preregistered
multi-lab study \citep{hagger2016multilab}.  A system retrieving only earlier
positive literature could still reconstruct it as a plausible research
program.  The governance concern is throughput.  One rediscovery is a review
failure; thousands of rediscoveries become a queueing problem for human
expertise.

\paragraph{Worked example: ego depletion as a corpus trap.}
Consider an AI scientist asked in 2026 to propose behavioural interventions
that improve self-control.  A retrieval step over the pre-replication
literature would find a coherent positive story: the original ego-depletion
claim, many follow-up experiments, and review articles treating the effect as
a live psychological construct.  The system would also find a different
evidential state if its corpus included preregistered failures and later
meta-scientific debate: the large multi-lab replication found little support
for the depletion effect \citep{hagger2016multilab}.  The failure mode arises
when the first evidence set is visible and the second is absent, buried, or
not retrieved.  The generator can then write a plausible proposal for
``AI-optimised ego-depletion interventions,'' and an automated evaluator can
reward it for novelty, social relevance, and citation support.  No component
needs to hallucinate.  The error is inherited from the corpus boundary.  This
is why the problem is governance-relevant: the correct fix is not only a
better prompt, but a retrieval substrate where failed replications and null
results are first-class evidence.

\paragraph{Ghost evidence accumulation.}
Multiple AI scientist systems draw on the same biased corpus and partially
validate the same false hypothesis.  Later systems cite earlier AI outputs as
if they were independent evidence.  The resulting citation network appears to
show convergence, even though the sources share the same upstream omission of
null results.  This is the hardest failure mode to audit because no individual
paper needs to be fraudulent.  Each output can be locally defensible given its
retrieved context, while the aggregate literature becomes increasingly
misleading.
This is no longer a purely hypothetical concern.  Population-level analyses
already find substantial LLM modification in scientific writing, especially in
computer-science preprints, and citation-manipulation work shows that
preprint-linked citation networks can be gamed or distorted at scale
\citep{liang2025quantifying,ibrahim2025citation}.  AI scientist outputs would
add another layer to this same networked evidence problem: machine-generated
claims can become inputs to later machine-generated claims before independent
experimental checks arrive.
The AI venue corpus itself already shows this failure mode in miniature:
GPTZero reported 100 confirmed hallucinated citations across 51 accepted
NeurIPS 2025 papers and more than 50 hallucinated citations in ICLR 2026
submissions \citep{shmatko2026neurips,esau2025iclr}.  Even if these errors
affect a small fraction of all citations, they demonstrate the path by which
machine-generated evidence can enter the literature and later appear as
ordinary scholarly context.

\paragraph{Replication laundering.}
An AI-generated claim is cited by another AI system as prior evidence, then
reappears as a confirmation.  The loop mimics replication but lacks
independent experimental contact.  It is detectable in principle by tracing
the citation graph, but only if provenance infrastructure exists.  In the
absence of such infrastructure, the social signal of citation count can be
mistaken for the scientific signal of independent replication.

\paragraph{Confidence miscalibration.}
The system reports high confidence on hypotheses with little or no empirical
replication support.  This is hard to detect because the output can look like
ordinary confident science.  The error becomes visible only when confidence is
compared with replication rates, retraction status, and downstream outcomes.
For AI scientist systems, calibration should therefore be evaluated at the
claim level and the corpus level, not only at the level of individual
question-answer accuracy.

\begin{figure*}[t]
\centering
\figwide{0.90\textwidth}{0.27\textheight}{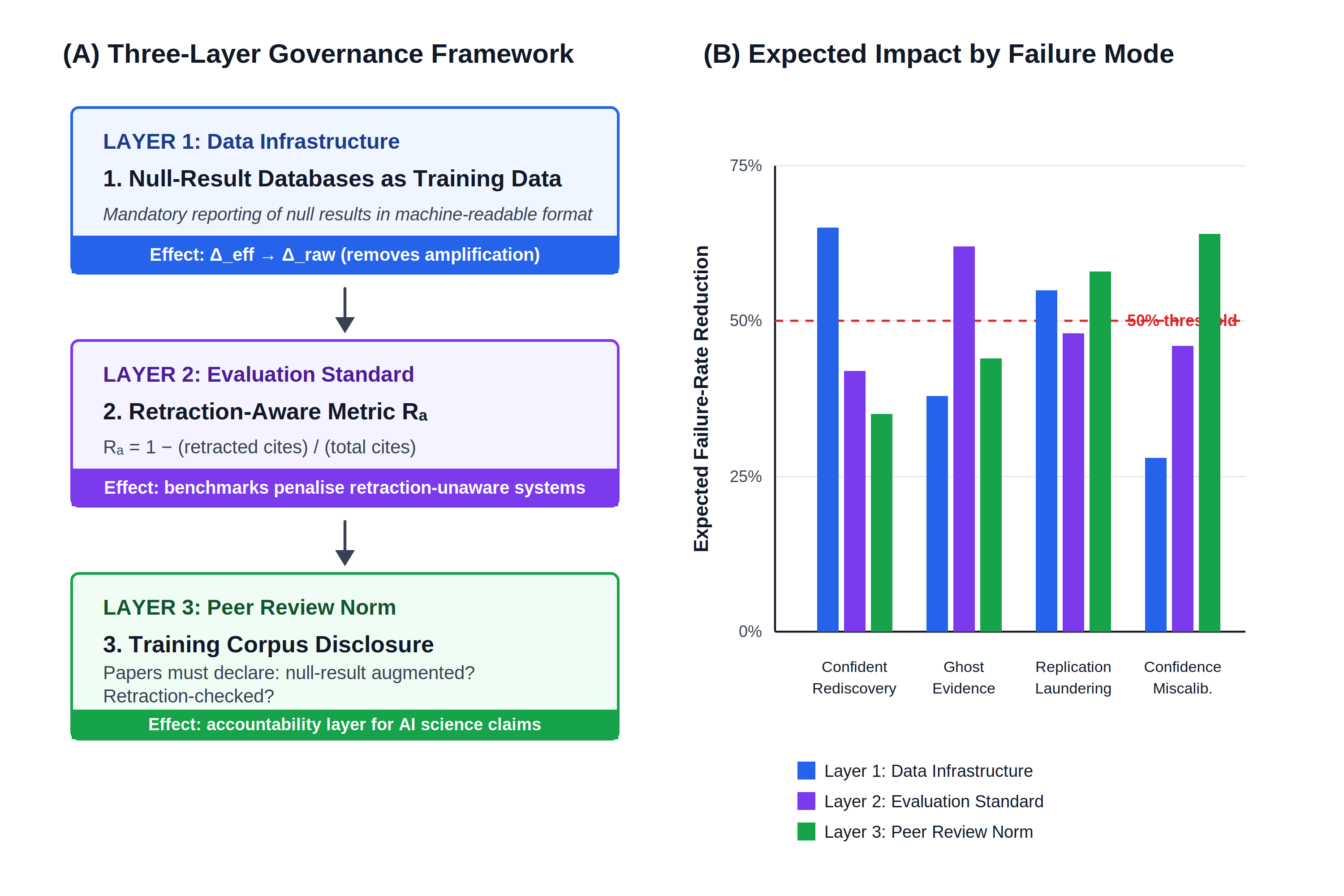}
\caption{\textbf{Three-layer governance framework.}
Layer 1 changes the retrieval substrate by making null results available.
Layer 2 changes evaluation incentives by penalising retraction-unaware
systems.  Layer 3 changes publication norms by requiring corpus disclosure.
The point is not to stop AI science, but to make its evidence base auditable.}
\label{fig:governance}
\end{figure*}

\section{Three Structural Interventions}
\label{sec:interventions}

The failure modes above share a common root: the training and retrieval corpus
does not faithfully represent the evidential state of science.  We propose
three interventions at different institutional layers.

\subsection{Null-Result Databases as Training Infrastructure}

AI scientist systems should be able to retrieve failed replications, negative
trials, and null results in machine-readable form.  We propose a structured
schema for null results, including hypothesis, protocol, outcome, effect size,
confidence interval, preregistration link, and provenance.  This extends the
logic of ClinicalTrials.gov, registered reports, and preregistration to
AI-readable scientific corpora
\citep{dickersin1992factors,zarin2011clinicaltrials,chambers2013registered,
nosek2015promoting}.  The expected effect
is to reduce $\Delta$ at the source.  Importantly, this is not merely a call
for more data.  It is a call for a different kind of data: records of what was
tried, what failed, and under what experimental conditions.  For an AI
scientist, a well-specified null result is not negative information.  It is a
map of where not to spend the next unit of experimental effort.
This proposal aligns with a recent 43-author consensus call in
\emph{PLOS Biology} for a values-based approach to surfacing null and negative
results \citep{curry2025ending}.  Their human-facing reforms and our
AI-facing infrastructure proposal are complementary: the same evidential gaps
that distort human science also distort the corpora AI scientists inherit.

\subsection{Retraction-Aware Evaluation}

Benchmarks for AI scientist systems should penalise reliance on retracted
literature.  Existing agent benchmarks such as MLE-bench help measure whether
systems can solve machine-learning engineering tasks, but governance-specific
benchmarks should also measure whether a system handles scientific provenance
correctly \citep{chan2024mlebench}.  We propose a simple retraction-aware
score:
\begin{equation}
  R_a =
  1 -
  \frac{|\{\text{uncontextualised retracted citations}\}|}
       {\max(1,|\{\text{claim-supporting citations}\}|)}.
  \label{eq:ra}
\end{equation}

$R_a$ is imperfect, but it is computable from Retraction Watch and Crossref
metadata \citep{retwatch2024}.
The numerator counts retracted work used as support without warning, not
neutral historical discussion of a retraction.  The denominator is restricted
to claim-supporting citations to reduce padding with irrelevant references;
outputs with no citations receive the fallback denominator $1$.

Its main virtue is incentive alignment:
systems that ignore retractions should not receive the same benchmark score as
systems that detect and contextualise them.  A stronger benchmark could also
distinguish between inappropriate positive citation, neutral historical
citation, and explicit discussion of why a retracted result should not be used.
As a baseline, a system citing broadly from PubMed or Semantic
Scholar may often obtain $R_a\approx0.998$ because retractions are rare
relative to the full literature.  That high baseline is not reassuring by
itself: $R_a$ becomes informative precisely on contested or fraud-prone
queries, where retracted and expression-of-concern papers are concentrated.
Empirical evaluations confirm the urgency.  In one study of 217 retracted or
otherwise concerning papers, ChatGPT-4o-mini produced 6,510 quality reports
without mentioning the retractions or reliability concerns in any report
\citep{thelwall2025chatgpt}. A separate study found that ChatGPT-4o, DeepSeek, and Grok used 84 of 93
retracted stem-cell articles in answers \citep{yao2025misuse}.  $R_a$
therefore measures a documented weakness in deployed literature-facing
systems, not a hypothetical failure mode.

\subsection{Training Corpus Disclosure}

Venues accepting AI-generated or AI-assisted scientific work should require a
training corpus card.  This follows the spirit of model cards, datasheets for
datasets, data statements, data nutrition labels, and broader documentation
standards for machine-learning artifacts
\citep{mitchell2019model,gebru2021datasheets,bender2018data,
holland2018dataset,raji2020closing}.  The card should disclose sources, null-result coverage,
retraction filtering, knowledge cutoff, and whether generated papers can enter
future training or retrieval corpora.  Corpus disclosure does not eliminate
bias, but it makes bias auditable. Instead of asking reviewers to infer whether a system used a biased
corpus, authors would be expected to state what corpus was used, how it was
filtered, and what categories of evidence were structurally absent.
\paragraph{What disclosure changes.}
A corpus card is not meant to certify that a system is unbiased.  Its purpose
is to make the evidential boundary visible.  Reviewers should be able to tell
whether the system had access to failed replications, whether retracted papers
were removed or only flagged, and whether AI-generated outputs were allowed to
re-enter retrieval as apparently independent evidence.  This changes corpus
quality from an implicit implementation detail into an auditable part of the
scientific claim. Null-result coverage should be operationalised at the claim-record level: the
numerator is the number of indexed claims whose outcome is negative, null,
failed, contradicted, or non-replicating, and the denominator is the total
number of indexed empirical claim records in the corpus.  Reporting this value
does not certify completeness, but it prevents null evidence from remaining an
invisible implementation detail.

\begin{center}
\begin{minipage}{\columnwidth}
\captionof{table}{\textbf{Example training corpus card} for an AI scientist system.}
\label{tab:corpus-card}
\vspace{2pt}
\centering
\small
\setlength{\tabcolsep}{4pt}
\renewcommand{\arraystretch}{1.12}
\begin{tabular}{p{0.35\columnwidth}p{0.56\columnwidth}}
\toprule
Field & Example entry \\
\midrule
System & AI Chemistry Scientist v1.0 \\
Sources & PubMed, arXiv, patents, internal ELN \\
Null-result coverage & 8\% explicit negative or failed assays \\
Retraction filtering & Yes; Crossref and Retraction Watch \\
Knowledge cutoff & Literature indexed through Jan. 2026 \\
Feedback loop & Generated reports excluded from retrieval until human audit \\
\bottomrule
\end{tabular}
\end{minipage}
\end{center}

\paragraph{Why institutional rather than purely technical?}
Better retrieval, calibration, and benchmark scores can reduce errors, but they
do not create missing null results or reveal whether a system trained on
retracted evidence. The missing layer is institutional: AI scientist papers
should be evaluated not only by what their agents produce, but by what evidence
their agents were allowed to see.

\section{Discussion}
\label{sec:discussion}

Publication bias was already harmful when humans moved slowly.  AI scientist
systems change the dynamics because generation can outrun correction: a single
mistaken paper is familiar, but a self-reinforcing stream of machine-generated
claims entering the corpus for future systems is a different problem. This is not an argument against AI science.  AI scientist systems could
accelerate discovery, improve reproducibility, and make exploration cheaper.
The claim is narrower: acceleration is not automatically epistemic progress.
If the input record omits null results, then speed can amplify omission. The estimates of $\Delta$ and $\alpha_i$ are approximate, and the taxonomy is
conceptual rather than validated.  Still, adjacent deployed systems already
show related mechanisms: ego-depletion retrieval illustrates confident
rediscovery, hallucinated citations show ghost evidence accumulation, and
retraction-blind evaluations show confidence miscalibration.  Replication
laundering remains supported structurally rather than directly. The next step is empirical calibration: benchmark corpora with known
null-result coverage, retrieval and generation tests over matched hypotheses,
and LLM-judge evaluations of positive versus null or mixed narratives under
controlled factual support.  These experiments would turn this position paper
into a benchmark program for responsible AI science.

\section{Conclusion}
\label{sec:conclusion}

The null result gap is not a property of language models.  It is a property of
the publication system that language models learn from.  AI scientist systems
can inherit that gap, amplify it through retrieval and evaluation, and return
old failures as new discoveries.  The response should be infrastructural:
index null results, evaluate retraction awareness, and disclose training
corpora.  None of these steps requires slowing scientific AI.  They make acceleration
more trustworthy.  The aim is not to make AI scientists cautious by default,
but to make their evidence substrate inspectable before their conclusions
become reusable scientific infrastructure.  The AI-for-science community builds
the systems, designs the benchmarks, and runs the venues. It can set these norms before biased automation becomes scientific common sense;
after that point, the field will not be accelerating discovery so much as
accelerating the recycling of its own uncorrected errors.

\section*{Impact Statement}

This paper identifies a governance risk for AI scientist systems:
corpus-induced bias propagated at machine speed.  The proposed interventions
are additive to existing scientific infrastructure and do not require
restricting responsible AI-science deployment.  Their purpose is to improve the
evidential substrate on which AI scientists operate, so acceleration does not
come at the expense of reliability. More broadly, the paper argues that scientific acceleration should be evaluated
not only by throughput, but by the quality, provenance, and corrigibility of
the evidence being accelerated.  The broader societal benefit is a scientific
automation stack that accelerates discovery while preserving the conditions
for correction.

\bibliography{example_paper}

@article{ioannidis2005most,
  title={Why most published research findings are false},
  author={Ioannidis, John P. A.},
  journal={PLOS Medicine}, volume={2}, number={8}, pages={e124}, year={2005},
  doi={10.1371/journal.pmed.0020124}
}

@article{fanelli2010positive,
  title={{``Positive''} results increase down the hierarchy of the sciences},
  author={Fanelli, Daniele},
  journal={PLOS ONE}, volume={5}, number={4}, pages={e10068}, year={2010},
  doi={10.1371/journal.pone.0010068}
}

@article{franco2014publication,
  title={Publication bias in the social sciences: Unlocking the file drawer},
  author={Franco, Annie and Malhotra, Neil and Simonovits, Gabor},
  journal={Science}, volume={345}, number={6203}, pages={1502--1505}, year={2014},
  doi={10.1126/science.1255484}
}

@article{osc2015estimating,
  title={Estimating the reproducibility of psychological science},
  author={{Open Science Collaboration}},
  journal={Science}, volume={349}, number={6251}, pages={aac4716}, year={2015},
  doi={10.1126/science.aac4716}
}

@article{begley2012drug,
  title={Drug development: Raise standards for preclinical cancer research},
  author={Begley, C. Glenn and Ellis, Lee M.},
  journal={Nature}, volume={483}, number={7391}, pages={531--533}, year={2012},
  doi={10.1038/483531a}
}

@article{errington2021investigating,
  title={Investigating the replicability of preclinical cancer biology},
  author={Errington, Timothy M. and Mathur, Maya and Soderberg, Courtney K. and Denis, Alexandria and Perfito, Nicole and Iorns, Elizabeth and Nosek, Brian A.},
  journal={eLife}, volume={10}, pages={e71601}, year={2021},
  doi={10.7554/eLife.71601}
}

@article{freedman2015economics,
  title={The economics of reproducibility in preclinical research},
  author={Freedman, Leonard P. and Cockburn, Iain M. and Simcoe, Timothy S.},
  journal={PLOS Biology}, volume={13}, number={6}, pages={e1002165}, year={2015},
  doi={10.1371/journal.pbio.1002165}
}

@article{steen2013retractions,
  title={Retractions in the medical literature: How many patients are put at risk by flawed research?},
  author={Steen, R. Grant and Casadevall, Arturo and Fang, Ferric C.},
  journal={Journal of Medical Ethics}, volume={39}, number={1}, pages={20--25}, year={2013},
  doi={10.1136/medethics-2012-100766}
}

@article{bar2021retracted,
  title={Retracted papers remain in the scientific literature},
  author={Bar-Ilan, Judit and Halevi, Gali},
  journal={Scientometrics}, volume={126}, number={7}, pages={6029--6045}, year={2021},
  doi={10.1007/s11192-021-03967-2}
}

@misc{retwatch2024,
  title={Retraction Watch Database},
  author={{Retraction Watch}},
  howpublished={\url{https://retractionwatch.com/}},
  year={2024},
  note={Accessed 2026}
}

@article{nosek2015promoting,
  title={Promoting an open research culture},
  author={Nosek, Brian A. and Alter, George and Banks, George C. and Borsboom, Denny and Bowman, Sara D. and Breckler, Steven J. and Buck, Stuart and Chambers, Christopher D. and Chin, Gilbert and Christensen, Garret and others},
  journal={Science}, volume={348}, number={6242}, pages={1422--1425}, year={2015},
  doi={10.1126/science.aab2374}
}

@article{chambers2013registered,
  title={Registered reports: A new publishing initiative at {Cortex}},
  author={Chambers, Christopher D.},
  journal={Cortex}, volume={49}, number={3}, pages={609--610}, year={2013},
  doi={10.1016/j.cortex.2012.12.016}
}

@article{hagger2016multilab,
  title={A multilab preregistered replication of the ego-depletion effect},
  author={Hagger, Martin S. and Chatzisarantis, Nikos L. D. and Alberts, Hugo and Anggono, Calvin O. and Batailler, Cedric and Birt, Angela R. and Brand, Ralf and Brandt, Mark J. and Brewer, Gene and Bruyneel, Sabrina and others},
  journal={Perspectives on Psychological Science}, volume={11}, number={4}, pages={546--573}, year={2016},
  doi={10.1177/1745691616652873}
}

@article{dickersin1992factors,
  title={Factors influencing publication of research results: Follow-up of applications submitted to two institutional review boards},
  author={Dickersin, Kay and Min, Yuan-I. and Meinert, Curtis L.},
  journal={JAMA}, volume={267}, number={3}, pages={374--378}, year={1992},
  doi={10.1001/jama.1992.03480030052036}
}

@article{zarin2011clinicaltrials,
  title={The {ClinicalTrials.gov} results database: Update and key issues},
  author={Zarin, Deborah A. and Tse, Tony and Williams, Rebecca J. and Califf, Robert M. and Ide, Nicholas C.},
  journal={New England Journal of Medicine}, volume={364}, number={9}, pages={852--860}, year={2011},
  doi={10.1056/NEJMsa1012065}
}

@misc{lu2024ai,
  title={The {AI} Scientist: Towards Fully Automated Open-Ended Scientific Discovery},
  author={Lu, Chris and Lu, Cong and Lange, Robert Tjarko and Foerster, Jakob and Clune, Jeff and Ha, David},
  journal={arXiv preprint arXiv:2408.06292},
  year={2024},
  doi={10.48550/arXiv.2408.06292}
}

@misc{gottweis2025ai,
  title={{AI} co-scientist: Accelerating scientific discovery with {AI}-powered research agents},
  author={Gottweis, Juraj and others},
  journal={arXiv preprint arXiv:2502.18864},
  year={2025}
}

@article{boiko2023autonomous,
  title={Autonomous chemical research with large language models},
  author={Boiko, Daniil A. and MacKnight, Robert and Kline, Ben and Gomes, Gabe},
  journal={Nature}, volume={624}, pages={570--578}, year={2023},
  doi={10.1038/s41586-023-06792-0}
}

@article{bran2024chemcrow,
  title={{ChemCrow}: Augmenting large-language models with chemistry tools},
  author={Bran, Andres M. and Cox, Sam and Schilter, Oliver and Baldassari, Carlo and White, Andrew D. and Schwaller, Philippe},
  journal={Nature Machine Intelligence}, volume={6}, pages={525--535}, year={2024},
  doi={10.1038/s42256-024-00832-8}
}

@article{swanson2024virtual,
  title={The virtual lab: {AI} agents design new {SARS-CoV-2} nanobodies with experimental validation},
  author={Swanson, Kyle and others},
  journal={bioRxiv}, year={2024},
  doi={10.1101/2024.11.11.623004}
}

@inproceedings{lewis2020retrieval,
  title={Retrieval-augmented generation for knowledge-intensive {NLP} tasks},
  author={Lewis, Patrick and Perez, Ethan and Piktus, Aleksandra and Petroni, Fabio and Karpukhin, Vladimir and Goyal, Naman and others},
  booktitle={Advances in Neural Information Processing Systems},
  volume={33}, pages={9459--9474}, year={2020}
}

@inproceedings{lin2022truthfulqa,
  title={Truthful{QA}: Measuring how models mimic human falsehoods},
  author={Lin, Stephanie and Hilton, Jacob and Evans, Owain},
  booktitle={Proceedings of the 60th Annual Meeting of the Association for Computational Linguistics},
  pages={3214--3252}, year={2022},
  doi={10.18653/v1/2022.acl-long.229}
}

@misc{si2024llm,
  title={{LLM}-generated literature reviews: Identifying challenges and proposing solutions},
  author={Si, Chenglei and Yang, Diyi and Hashimoto, Tatsunori},
  journal={arXiv preprint arXiv:2403.04726},
  year={2024}
}

@inproceedings{zheng2024judging,
  title={Judging {LLM}-as-a-judge with {MT-Bench} and chatbot arena},
  author={Zheng, Lianmin and Chiang, Wei-Lin and Sheng, Ying and Zhuang, Siyuan and Wu, Zhanghao and Zhuang, Yonghao and Lin, Zi and Li, Zhuohan and Li, Dacheng and Xing, Eric P. and others},
  booktitle={Advances in Neural Information Processing Systems},
  volume={36}, year={2024}
}

@misc{li2024llmjudge,
  title={A survey on {LLM}-as-a-judge},
  author={Li, Junlong and others},
  journal={arXiv preprint arXiv:2411.15594},
  year={2024}
}

@misc{chen2024humans,
  title={Are {LLMs} good reviewers? Measuring human and {GPT}-4 agreement in peer review},
  author={Chen, Jiaao and others},
  journal={arXiv preprint arXiv:2412.01099},
  year={2024}
}

@inproceedings{mitchell2019model,
  title={Model cards for model reporting},
  author={Mitchell, Margaret and Wu, Simone and Zaldivar, Andrew and Barnes, Parker and Vasserman, Lucy and Hutchinson, Ben and Spitzer, Elena and Raji, Inioluwa Deborah and Gebru, Timnit},
  booktitle={Proceedings of the Conference on Fairness, Accountability, and Transparency},
  pages={220--229}, year={2019},
  doi={10.1145/3287560.3287596}
}

@article{gebru2021datasheets,
  title={Datasheets for datasets},
  author={Gebru, Timnit and Morgenstern, Jamie and Vecchione, Briana and Vaughan, Jennifer Wortman and Wallach, Hanna and Daum{\'e} III, Hal and Crawford, Kate},
  journal={Communications of the ACM}, volume={64}, number={12}, pages={86--92}, year={2021},
  doi={10.1145/3458723}
}

@article{bender2018data,
  title={Data statements for natural language processing: Toward mitigating system bias and enabling better science},
  author={Bender, Emily M. and Friedman, Batya},
  journal={Transactions of the Association for Computational Linguistics},
  volume={6}, pages={587--604}, year={2018},
  doi={10.1162/tacl{\_}a{\_}00041}
}

@misc{chan2024mlebench,
  title={{MLE}-bench: Evaluating machine learning agents on machine learning engineering},
  author={Chan, Jun Shern and Pieler, Niko and Jao, Justin and Peetoom, Jasper and Mullen-Schultz, Rachel and others},
  journal={arXiv preprint arXiv:2410.07095},
  year={2024}
}

@misc{anthropic2025scienceprogram,
  title={Introducing {Anthropic}'s {AI} for Science Program},
  author={{Anthropic}},
  howpublished={\url{https://www.anthropic.com/news/ai-for-science-program}},
  year={2025},
  note={Anthropic News, accessed 2026}
}

@misc{anthropic2025aijam,
  title={{Anthropic} partners with {U.S.} National Labs for first 1,000 Scientist {AI} Jam},
  author={{Anthropic}},
  year={2025},
  howpublished={\url{https://www.anthropic.com/news/anthropic-partners-with-u-s-national-labs-for-first-1-000-scientist-ai-jam}},
  note={Anthropic News, accessed 2026}
}

@misc{anthropic2026science,
  title={Science},
  author={{Anthropic}},
  year={2026},
  howpublished={\url{https://www.anthropic.com/science}},
  note={Anthropic Science page, accessed 2026}
}

@article{simmons2011false,
  title={False-positive psychology: Undisclosed flexibility in data collection and analysis allows presenting anything as significant},
  author={Simmons, Joseph P. and Nelson, Leif D. and Simonsohn, Uri},
  journal={Psychological Science}, volume={22}, number={11}, pages={1359--1366}, year={2011},
  doi={10.1177/0956797611417632}
}

@article{gelman2013garden,
  title={The garden of forking paths: Why multiple comparisons can be a problem, even when there is no fishing expedition or p-hacking and the research hypothesis was posited ahead of time},
  author={Gelman, Andrew and Loken, Eric},
  journal={Department of Statistics, Columbia University},
  year={2013}
}

@article{sterne2001funnel,
  title={Funnel plots for detecting bias in meta-analysis: Guidelines on choice of axis},
  author={Sterne, Jonathan A. C. and Egger, Matthias},
  journal={Journal of Clinical Epidemiology}, volume={54}, number={10}, pages={1046--1055}, year={2001},
  doi={10.1016/S0895-4356(01)00377-8}
}

@article{byrne2022paper,
  title={The fight against fake-paper factories that churn out sham science},
  author={Byrne, Jennifer A.},
  journal={Nature}, volume={603}, number={7901}, pages={370--371}, year={2022},
  doi={10.1038/d41586-022-00733-5}
}

@inproceedings{maynez2020faithfulness,
  title={On faithfulness and factuality in abstractive summarization},
  author={Maynez, Joshua and Narayan, Shashi and Bohnet, Bernd and McDonald, Ryan},
  booktitle={Proceedings of the 58th Annual Meeting of the Association for Computational Linguistics},
  pages={1906--1919}, year={2020},
  doi={10.18653/v1/2020.acl-main.173}
}

@article{ji2023survey,
  title={Survey of hallucination in natural language generation},
  author={Ji, Ziwei and Lee, Nayeon and Frieske, Rita and Yu, Tiezheng and Su, Dan and Xu, Yan and Ishii, Etsuko and Bang, Ye Jin and Madotto, Andrea and Fung, Pascale},
  journal={ACM Computing Surveys}, volume={55}, number={12}, pages={1--38}, year={2023},
  doi={10.1145/3571730}
}

@article{gao2023retrievalaugmented,
  title={Retrieval-augmented generation for large language models: A survey},
  author={Gao, Yunfan and Xiong, Yun and Gao, Xinyu and Jia, Kangxiang and Pan, Jinliu and Bi, Yuxi and Dai, Yi and Sun, Jiawei and Wang, Meng and Wang, Haofen},
  journal={arXiv preprint arXiv:2312.10997},
  year={2023}
}

@article{wang2023selfconsistency,
  title={Self-consistency improves chain of thought reasoning in language models},
  author={Wang, Xuezhi and Wei, Jason and Schuurmans, Dale and Le, Quoc and Chi, Ed and Narang, Sharan and Chowdhery, Aakanksha and Zhou, Denny},
  journal={International Conference on Learning Representations},
  year={2023}
}

@article{holland2018dataset,
  title={The dataset nutrition label: A framework to drive higher data quality standards},
  author={Holland, Sarah and Hosny, Ahmed and Newman, Sarah and Joseph, Joshua and Chmielinski, Kasia},
  journal={arXiv preprint arXiv:1805.03677},
  year={2018}
}

@inproceedings{raji2020closing,
  title={Closing the {AI} accountability gap: Defining an end-to-end framework for internal algorithmic auditing},
  author={Raji, Inioluwa Deborah and Smart, Andrew and White, Rebecca N. and Mitchell, Margaret and Gebru, Timnit and Hutchinson, Ben and Smith-Loud, Jamila and Theron, Daniel and Barnes, Parker},
  booktitle={Proceedings of the Conference on Fairness, Accountability, and Transparency},
  pages={33--44}, year={2020},
  doi={10.1145/3351095.3372873}
}

@article{duyx2019abstract,
  title={The strong focus on positive results in abstracts may cause bias in systematic reviews: A case study on abstract reporting bias},
  author={Duyx, Bram and Swaen, Gerard M. H. and Urlings, Miriam J. E. and Bouter, Lex M. and Zeegers, Maurice P.},
  journal={Systematic Reviews}, volume={8}, number={1}, pages={217}, year={2019},
  doi={10.1186/s13643-019-1082-9}
}

@article{dwan2008systematic,
  title={Systematic review of the empirical evidence of study publication bias and outcome reporting bias},
  author={Dwan, Kerry and Altman, Douglas G. and Arnaiz, Juan A. and Bloom, Jill and Chan, An-Wen and Cronin, Eugenia and Decullier, Evelyne and Easterbrook, Philippa J. and Von Elm, Erik and Gamble, Carrol and Ghersi, Davina and Ioannidis, John P. A. and Simes, John and Williamson, Paula R.},
  journal={PLOS ONE}, volume={3}, number={8}, pages={e3081}, year={2008},
  doi={10.1371/journal.pone.0003081}
}

@article{jumper2021highly,
  title={Highly accurate protein structure prediction with {AlphaFold}},
  author={Jumper, John and Evans, Richard and Pritzel, Alexander and Green, Tim and Figurnov, Michael and Ronneberger, Olaf and Tunyasuvunakool, Kathryn and Bates, Russ and Zidek, Augustin and Potapenko, Anna and others},
  journal={Nature}, volume={596}, number={7873}, pages={583--589}, year={2021},
  doi={10.1038/s41586-021-03819-2}
}

@article{liang2025quantifying,
  title={Quantifying large language model usage in scientific papers},
  author={Liang, Weixin and Zhang, Yaohui and Wu, Zhengxuan and Lepp, Haley and Ji, Wenlong and Zhao, Xuandong and Cao, Hancheng and Liu, Sheng and He, Siyu and Huang, Zhi and Yang, Diyi and Potts, Christopher and Manning, Christopher D. and Zou, James},
  journal={Nature Human Behaviour}, volume={9}, pages={2599--2609}, year={2025},
  doi={10.1038/s41562-025-02273-8}
}

@article{ibrahim2025citation,
  title={Citation manipulation through citation mills and pre-print servers},
  author={Ibrahim, Hazem and Liu, Fengyuan and Zaki, Yasir and Rahwan, Talal and others},
  journal={Scientific Reports}, volume={15}, pages={5480}, year={2025},
  doi={10.1038/s41598-025-88709-7}
}

@article{peters2025generalization,
  title={Generalization bias in large language model summarization of scientific research},
  author={Peters, Uwe and Chin-Yee, Benjamin},
  journal={Royal Society Open Science}, volume={12}, number={4}, pages={241776}, year={2025},
  doi={10.1098/rsos.241776}
}

@article{alkaissi2023artificial,
  title={Artificial hallucinations in {ChatGPT}: Implications in scientific writing},
  author={Alkaissi, Hussam and McFarlane, Samy I.},
  journal={Cureus}, volume={15}, number={2}, pages={e35179}, year={2023},
  doi={10.7759/cureus.35179}
}

@article{sidhu2026trust,
  title={Trust, truth and transparency: Analysing the references underpinning {AI}-generated surgical information},
  author={Sidhu, Rickvir S. and Selvamogan, Arrane and Boddy, Amy M.},
  journal={Annals of The Royal College of Surgeons of England},
  year={2026},
  doi={10.1308/rcsann.2026.0021}
}

@article{shumailov2024collapse,
  title={{AI} models collapse when trained on recursively generated data},
  author={Shumailov, Ilia and Shumaylov, Zakhar and Zhao, Yiren and Papernot, Nicolas and Anderson, Ross and Gal, Yarin},
  journal={Nature}, volume={631}, pages={755--759}, year={2024},
  doi={10.1038/s41586-024-07566-y}
}

@misc{shmatko2026neurips,
  title={{GPTZero} finds 100 new hallucinations in {NeurIPS} 2025 accepted papers},
  author={Shmatko, Nazar and Adam, Alex and Esau, Paul},
  howpublished={\url{https://gptzero.me/news/neurips/}},
  year={2026},
  note={Accessed 2026}
}

@misc{esau2025iclr,
  title={{GPTZero} finds over 50 new hallucinations in {ICLR} 2026 submissions},
  author={Esau, Paul and Shmatko, Nazar and Adam, Alex},
  howpublished={\url{https://gptzero.me/news/iclr-2026/}},
  year={2025},
  note={Accessed 2026}
}

@article{curry2025ending,
  title={Ending publication bias: A values-based approach to surface null and negative results},
  author={Curry, Stephen and Mercado-Lara, Eunice and Arechavala-Gomeza, Virginia and Begley, C. Glenn and Bernard, Christophe and Bernard, Ren{\'e} and Bertuzzi, Stefano and Bhalla, Needhi and Bowers, Dawn and Brod, Samuel and others},
  journal={PLOS Biology}, volume={23}, number={9}, pages={e3003368}, year={2025},
  doi={10.1371/journal.pbio.3003368}
}

@article{thelwall2025chatgpt,
  title={Does {ChatGPT} ignore article retractions and other reliability concerns?},
  author={Thelwall, Mike and Lehtisaari, Marianna and Katsirea, Irini and Holmberg, Kim and Zheng, Er-Te},
  journal={Learned Publishing}, volume={38}, number={4}, year={2025},
  doi={10.1002/leap.2018}
}

@article{yao2025misuse,
  title={{AI} misuse of retracted literature: A comparative study of {ChatGPT4o}, {DeepSeek}, and {Grok} 3 in stem cell research},
  author={Yao, Lan and Gu, Tianshu and Li, Xuexin and Jiao, Yan and Li, Minghui and Li, Yulan and Graff, J. Carolyn and Gu, Weikuan},
  journal={The Science of Nature}, volume={112}, pages={85}, year={2025},
  doi={10.1007/s00114-025-02036-5}
}
\bibliographystyle{icml2026}

\end{document}